# Stripes in Microscopic Theory of High-T$_c$ Superconductivity

Daniel C. Mattis

Department of Physics, University of Utah

115 S. 1400 E.  #201

Salt Lake City, UT 84112-0830

Abstract:

We examine a realistic 3-band model, finding it capable of exhibiting the *d*-wave pairing characteristic of CuO$_2$-based high-$T_c$ superconductors, but only in the presence of symmetry-lowering charge-density "stripes" aligned along (1,0) axes, preferably of diameter 2 cell-widths. The present theoretical treatment has no adjustable parameters (the unit of energy $t^2/V$ factors out of both kinetic and potential energy terms,) does not require fine-tuning, and is in qualitative accord with experiment.

*INTRODUCTION.* The present paper reëxamines the exact causes of high-$T_c$ superconductivity, a topic that has been without resolution for almost 20 years. There is a general belief, to which we also subscribe, that the electron-phonon interaction is incapable of explaining superconductivity in the 100 K range or higher. Here we propose a theory based on linear combinations of copper and oxygen orbitals. This 3-band model does the job – but only if the added charges are distributed, at least in part, inhomogeneously, in alternating lines or *stripes*.

The mean-field analysis adduced here in support of these conclusions does *not* require "fine-tuning" as there are no adjustable parameters. The initial assumptions are merely that the potential energy parameter $|V|$ be large compared to the "hopping" matrix element $t$, and that a repulsive two-body potential $U>>V$ be larger still. A single parameter $g_1 \equiv t^2/V$ sets the scale of energies; because it multiplies the *entire* pairing Hamiltonian, the results are universal.

*HUBBARD MODEL.* We *cannot* use Hubbard's familiar model of interacting electrons[1] to study high-$T_c$ superconductivity, although it is generally considered a reasonable facsimile of the electron gas, insofar as it is based on sound atomic considerations.[2] Although it has occasionally been claimed that it also explains virtually all aspects of high-$T_c$ superconductivity,[3] such claims have not been universally supported.[4,5,6,7,8]

Our reason for dropping the ordinary Hubbard model is practical. Even when it is augmented by stripes and solved similarly to the 3-band model, the "pairing" term, proportional to $t^2/U$, is *much* smaller than the motional (kinetic energy) terms, which are proportional to $t$. Thus, computationally, it fails to yield a sturdy or reliable solution.

*THREE-BAND MODEL.* Let us start by describing the geometry that drives our model.[9] Although well known, it bears repeating: ligand oxygen ions $O^{2-}$ in the $CuO_2$ lattice decorate what is basically a 2D *sq* lattice of copper $Cu^{2+}$ ions. The unit cell contains 3 inequivalent sites, a geometry that favors a 3-band model for the holes. But even in this more complex situation, we find that a *charged pattern* that lowers $C_4$ symmetry to $C_2$ is required. We shall see that such a pattern is required to stabilize the *d*-wave pairing commonly observed in high-$T_c$ superconductors.

Label the intercalated oxygens on the horizontal lines by $a_R$ and on verticals by $b_{R'}$. To mimic an assumed orbital energy mismatch, a potential $-V$ is assigned to *holes* on *copper sites* and $+V$ on *oxygen* sites. A potential $-V$ promotes valency $Cu^{2+}$ over $Cu^{1+}$. To inhibit further (unwanted) oxidation to $Cu^{3+}$, a Hubbard-like two-body potential $U$ must be introduced. It repulses 2 particles[10] on the same copper site. Because there are so few additional particles added to the oxygen sites (superconductivity is optimized at 1 added particle for every 8 oxygen sites,) the random statistical tendency to $O^{0-}$ is tiny. Therefore any error we commit in neglecting two-body repulsions on the oxygen sites is negligible.

Initially $H$ takes the form:

$$H = -t\sum_{R,\sigma}\{c^\dagger_{R,\sigma}(a_{R+(\frac{1}{2},0),\sigma} + a_{R+(\frac{-1}{2},0),\sigma} + b_{R+(0,\frac{1}{2}),\sigma} + b_{R+(0,\frac{-1}{2}),\sigma}) + H.c.\} + V\sum_{R,\sigma}\{a^\dagger_{R+(\frac{1}{2},0),\sigma}a_{R+(\frac{1}{2},0),\sigma}$$
$$+ b^\dagger_{R+(0,\frac{1}{2}),\sigma}b_{R+(0,\frac{1}{2}),\sigma} - c^\dagger_{R,\sigma}c_{R,\sigma}\} + U^*\sum_R c^\dagger_{R,\uparrow}c_{R,\uparrow}c^\dagger_{R,\downarrow}c_{R,\downarrow} \quad (1)$$

where the last term in this Hamiltonian is a "projection" operator. Once $U$ exceeds $2V$ it no longer matters how large it really is. That is is why, in this introductory version of the theory, we immediately proceed to the strong-coupling limit of $U$ (using $U^* \equiv \infty$.)

Several exact transformations simplify the model greatly. First, a Fourier transform:

$$a_{R+(\frac{1}{2},0),\sigma} + a_{R+(\frac{-1}{2},0),\sigma} \equiv \frac{2}{\sqrt{N}} \sum_{k \subseteq BZ} a(k,\sigma) \cos\frac{k_x}{2} e^{ik\cdot R}$$

$$(b_{R+(0,\frac{1}{2}),\sigma} + b_{R+(0,\frac{-1}{2}),\sigma}) \equiv \frac{2}{\sqrt{N}} \sum_{k \subseteq BZ} b(k,\sigma) \cos\frac{k_y}{2} e^{ik\cdot R} \quad (2)$$

defines the bare band states (*BZ* indicates the first Brillouin Zone of the *sq* lattice.) Then,

$$\alpha(k,\sigma) \equiv \frac{a(k,\sigma)\cos\frac{k_x}{2} + b(k,\sigma)\cos\frac{k_y}{2}}{\sqrt{\cos^2\frac{k_x}{2} + \cos^2\frac{k_y}{2}}} \text{ and } \beta(k,\sigma) \equiv \frac{a(k,\sigma)\cos\frac{k_y}{2} - b(k,\sigma)\cos\frac{k_x}{2}}{\sqrt{\cos^2\frac{k_x}{2} + \cos^2\frac{k_y}{2}}} \quad (3)$$

are the new Bloch operators. Their associated Wannier operators are taken to be:

$$\alpha(k,\sigma) \equiv \frac{1}{\sqrt{N}} \sum_{R'} \alpha_{R',\sigma} e^{-ik\cdot R'}, \text{ i.e., } \alpha_{R',\sigma} \equiv \frac{1}{\sqrt{N}} \sum_{k \subseteq BZ} \alpha(k,\sigma) e^{ik\cdot R'} \quad (4)$$

and similarly for the $\beta$'s. The strong-coupling *H* now assumes the form:

$$H = -t \sum_{R,R'} \sum_{\sigma} G(R-R')\{c^\dagger_{R,\sigma}\alpha_{R',\sigma} + H.c.\} + V \sum_{R}\sum_{\sigma}\{\alpha^\dagger_{R,\sigma}\alpha_{R,\sigma} + \beta^\dagger_{R,\sigma}\beta_{R,\sigma} - c^\dagger_{R,\sigma}c_{R,\sigma}\}$$
$$+ U * \sum_{R} c^\dagger_{R,\uparrow} c_{R,\uparrow} c^\dagger_{R,\downarrow} c_{R,\downarrow} \quad . \quad (5)$$

It is unitarily equivalent to (1). The function *G(R)* and its Fourier transform *g(k)*, are:

$$G(R) \equiv \frac{1}{N}\sum_{k\subseteq BZ} e^{ik\cdot R} g(k_x, k_y) = \frac{1}{N}\sum_{k\subseteq BZ} e^{ik\cdot R}\left(2 \times \sqrt{\cos^2\frac{k_x}{2} + \cos^2\frac{k_y}{2}}\right)$$

$$= \frac{2}{\pi^2}\int_0^\pi dk_x \cos nk_x \int_0^\pi dk_y \cos mk_y \left(\sqrt{\cos^2\frac{k_x}{2} + \cos^2\frac{k_y}{2}}\right) \quad (6)$$

Values of *G(R)* are listed to 4 decimal places in Table I, at several small values of *R*.[9]

**TABLE I**

Values of $G(n,m) = G(\pm m, \pm n)$, where $G(0,0) = +1.9162$ and $g(k) = 2 \times \sqrt{\cos^2 \frac{k_x}{2} + \cos^2 \frac{k_y}{2}}$

| n,m → | 0,1 | 0,2 | 1,1 | 0,3 | 1,2 | 0,4 | 1,3 | 2,2 |
|---|---|---|---|---|---|---|---|---|
| G(n,m)→ | +0.2802 | –0.0275 | –0.4701 | 0.0071 | 0.0137 | –0.0027 | –0.0052 | –0.0065 |

*STATES of THE INTRASITE HAMILTONIAN.* Write $H=H_0+H'$, where $H_0$ is the *intra*site Hamiltonian $\propto G(0)$. The *inter*site $H'$ appears to be almost an order of magnitude smaller – insomuch as it is parametrized by $G(R)$'s at $R \neq 0$. Then this "larger" intrasite part is,

$$H_0 = -t \sum_R \sum_\sigma G(0) \{c^\dagger_{R,\sigma} \alpha_{R,\sigma} + H.c.\} + V \sum_R \sum_\sigma \{\alpha^\dagger_{R,\sigma} \alpha_{R,\sigma} + \beta^\dagger_{R,\sigma} \beta_{R,\sigma} - c^\dagger_{R,\sigma} c_{R,\sigma}\}$$
$$+ U^* \sum_R c^\dagger_{R,\uparrow} c_{R,\uparrow} c^\dagger_{R,\downarrow} c_{R,\downarrow} \tag{7}$$

Because $H_0 \equiv \sum_R H_0(R)$, its eigenstates are products, $|\Psi\rangle_0 = \prod_R |\Phi_{\alpha(R)}\rangle$. They comprise a complete basis set in which to expand the eigenstates of the total $H$.

At any fixed $R$ the projection operator in $U^*$ allows only certain states. As for the $\beta$'s, they are, and will forever remain, disconnected from all other operators.[11] The individual eigenstates at fixed $R$ are evaluated next, ranked according to ascending particle occupancy. We retain only those states having lowest possible energy eigenvalues in $H_0$ and the excited states to which they connect strongly *via* the "perturbation" $H'$.

<u>0 occupancy</u>. The site "vacuum" is denoted "vacancy" or |0). Its eigenvalue is $E_0=0$.

<u>1 occupancy</u>. There are 2 states of a single particle, one for each spin orientation $\sigma$. The 2×2 $H_0$ matrix at fixed $R$ is constructed with the help of $H_0(R)$ above. The 1-particle eigenoperator $d_\sigma^\dagger(R)|0)$ consists of the following linear combination,

$$\cos\phi\, c_{R,\sigma}^\dagger |0> + \sin\phi\, \alpha_{R,\sigma}^\dagger |0> \equiv d_\sigma^\dagger(R)|0) \tag{8}$$

Its energy eigenvalue is $E_1=\lambda_1$, where $\lambda$ is lower of the two solutions of the secular equation. When expanded in powers of $t/V$, it is:

$$\lambda_1 \approx -\{V + \frac{(tG(0))^2}{2V} - \frac{(tG(0))^4}{8V^3} + \ldots\} \text{ where, in (8),} \quad \tan\phi\, ;\, \frac{tG(0)}{2V} \quad . \tag{9}$$

This defines the ground state single occupancy spin-doublet. Further, define the corresponding basis states |1) and |2) at each $R$ by: $d_\uparrow^\dagger(R)|0) \equiv |1)$ and $d_\downarrow^\dagger(R)|0) \equiv |2)$. *Excited* states (belonging to eigenvalue $\lambda=+\{V+\ldots\}$ in Eq.(9)) are discarded, as are the $\beta$ states, the energy of which also $= +V$.

<u>Double Occupancy: spin singlets</u>. The lowest eigenstate in this subspace, denoted |3), is,

$$\frac{\cos\vartheta}{\sqrt{2}}(c_{R,\uparrow}^\dagger \alpha_{R,\downarrow}^\dagger - c_{R,\downarrow}^\dagger \alpha_{R,\uparrow}^\dagger)|0> + \sin\vartheta\, \alpha_{R,\uparrow}^\dagger \alpha_{R,\downarrow}^\dagger |0> \equiv D_R^\dagger |0) \equiv |3). \tag{10}$$

The above also serves to define "double occupancy" paulion operators $D$. The secular equation for the energy yields two solutions. With $g_1 \equiv t^2/V$ and $V >> t$, to leading order in $g_1$ the lower one has energy:

$$\lambda_2 \approx -g_1 G^2(0) \quad . \tag{11}$$

It is parametrized by an angle $\vartheta$. One easily finds: $\tan\vartheta \approx \frac{tG(0)}{V\sqrt{2}}$. The upper solution is discarded, as its energy exceeds $2V$.

Double Occupancy: triplets. There are 3 (i.e., *triplet*!) states, all with $\lambda=0$ precisely, which is neither low nor high in energy. In the interest of simplicity we *also* omit these states from further consideration here. In a separate paper under preparation,[12] their involvement and the new channel they open up is made clear. However, because triplets belong to a different symmetry the present results in the singlet channel remain unaltered.

Triple Occupancy. There are 2 such states of energy $\lambda=+V$. Because *in principle* two $D$'s could annihilate to form two doublets (consisting of 1 electron at one site and 3 at the other,) one should also introduce the corresponding operator $T^\dagger_{R,\sigma}|0) \equiv c^\dagger_{R,\sigma}\alpha^\dagger_{R,\uparrow}\alpha^\dagger_{R,\downarrow}|0>$, which has energy $+V$. Now, consider the reaction $D^\dagger_1 + D^\dagger_2 \rightleftharpoons T^\dagger_{1,\sigma} + d^\dagger_{2,-\sigma}$, which can proceed *iff* the energy of the final state $\leq$ than the initial, i.e. if $V-(V+\frac{(tG(0))^2}{2V}) \leq 2\lambda_2$.

This requires $-\frac{1}{2} g_1 \leq -2g_1$, an inequality that can *never* be satisfied. Hence the $D$'s are energetically stable against this process and triple-occupancy states are justifiably, and consistently, ignored. *Quadruple* occupancy is, of course, always forbidden by $U^*= +\infty$.

Table II summarizes matrix elements of operators in the reduced basis set, $|0) - |3)$: a matrix element $(i|op|j)$ is listed on a row labeled by that *op*, in the column marked $(i,j)$.

TABLE II:   SOME NONTRIVIAL MATRIX ELEMENTS

(in a fixed cell at a fixed *R*)

| Op ⇓ | (0,1) | (0,2) | (0,3) | (1,1) | (1,2) | (1,3) | (2,2) | (2,3) | (3,3) |
|---|---|---|---|---|---|---|---|---|---|
| $\alpha_\uparrow$ | $\sin\phi$ | 0 | 0 | 0 | 0 | 0 | 0 | F1 | 0 |
| $\alpha_\downarrow$ | 0 | $\sin\phi$ | 0 | 0 | 0 | $-F1$ | 0 | 0 | 0 |
| $c_\uparrow$ | $\cos\phi$ | 0 | 0 | 0 | 0 | 0 | 0 | F2 | 0 |
| $c_\downarrow$ | 0 | $\cos\phi$ | 0 | 0 | 0 | $-F2$ | 0 | 0 | 0 |
| $\alpha_\uparrow^\dagger \alpha_\uparrow$ | 0 | 0 | 0 | $\sin^2\phi$ | 0 | 0 | 0 | 0 | ½(1+$\sin^2\vartheta$) |
| $\alpha_\downarrow^\dagger \alpha_\downarrow$ | 0 | 0 | 0 | 0 | 0 | 0 | $\sin^2\phi$ | 0 | ½(1+$\sin^2\vartheta$) |
| $\alpha_\uparrow^\dagger \alpha_\downarrow$ | 0 | 0 | 0 | 0 | $\sin^2\phi$ | 0 | 0 | 0 | 0 |
| $c_\uparrow^\dagger c_\uparrow$ | 0 | 0 | 0 | $\cos^2\phi$ | 0 | 0 | 0 | 0 | ½(1−$\sin^2\vartheta$) |
| $c_\downarrow^\dagger c_\downarrow$ | 0 | 0 | 0 | 0 | 0 | 0 | $\cos^2\phi$ | 0 | ½(1−$\sin^2\vartheta$) |
| $c_\uparrow^\dagger c_\downarrow$ | 0 | 0 | 0 | 0 | $\cos^2\phi$ | 0 | 0 | 0 | 0 |

$$F1 = \frac{1}{\sqrt{2}}\cos\phi\cos\vartheta + \sin\phi\sin\vartheta, \quad F2 = \frac{1}{\sqrt{2}}\sin\phi\cos\vartheta$$

*INTERSITE DYNAMICS.* We turn to *H'*. *Given* a number of particles *exceeding* 1 particle *per* site on average, each vacancy permitted to exist at some *R* requires that at some other *R'*, a *d* is promoted →*D*. The net extra cost is thus +*V* per vacancy. Therefore vacancies must be disallowed at all concentrations exceeding 1 particle *per* site.

*Charge transfer*, a process connecting initially degenerate states, occurs by interchange of a 2-particle site at *R* with a 1-particle site at *R'*. It is the leading term in *H'*.

In terms of the bare operators, it is:

$$H' = -\frac{t}{2}\sum_{R,R'\neq R}\sum_\sigma G(R-R')\{c^\dagger_{R,\sigma}\alpha_{R',\sigma} + \alpha^\dagger_{R,\sigma}c_{R',\sigma} + c^\dagger_{R',\sigma}\alpha_{R,\sigma} + \alpha^\dagger_{R',\sigma}c_{R,\sigma}\}$$
$$+ U*\sum_R c^\dagger_{R,\uparrow}c_{R,\uparrow}c^\dagger_{R,\downarrow}c_{R,\downarrow} \qquad (13)$$

This now has to be rewritten in terms of the composite-fermion $d$ and composite-paulion $D$ using Table II and angles $\vartheta$ and $\phi$.

For each bond $(R,R')$ the *form* of the charge-transfer operator is $G(R-R')C(R,R')D_R^\dagger d_{R,\sigma} d_{R',\sigma}^\dagger D_{R'}$. Using Table II we can easily calculate $C(R,R')$. Thus, $c_{R,\sigma}^\dagger \Rightarrow \pm(F2)D_R^\dagger d_{R,-\sigma}$ and $\alpha_{R',\sigma} \Rightarrow \pm(F1)d_{R',-\sigma}^\dagger D_{R'}$. Their product yields $C(R,R') = (F1)(F2) = \frac{1}{\sqrt{2}}\sin\phi\cos\vartheta(\frac{1}{\sqrt{2}}\cos\phi\cos\vartheta + \sin\phi\sin\vartheta) \xrightarrow[V \gg t]{} \frac{tG(0)}{4V}$. Thus, the motional energy of the dressed particles is governed by:

$$KE = -\frac{g_1 G(0)}{4}\sum_R \sum_{R' \neq R} \sum_\sigma G(R-R')D_R^\dagger d_{R,\sigma} d_{R',\sigma}^\dagger D_{R'}(D_R D_R^\dagger). \tag{14}$$

It gives rise to a "band structure" for composite particles that is exact, up to corrections $O(t/V)^2$. A redundant factor $(D_R D_R^\dagger)$ has been inserted to *ensure* that the "target" site $R$ is unoccupied by a $D$. Moreover, for current to flow, a *D must be present* at the initial site $R'$. It follows that if there were no $D$'s present, e.g., if exactly 1 particle *per* site is stipulated, (14) cannot carry current. This is in perfect accord with observation and has no connection with "nesting", etc.[13] For more than 1 particle *per* site the *KE* operator permits the excess charges to flow and to establish mutual correlations.

Next, the *in*elastic part of the same operator connects two singly occupied sites linked by $G(R–R')$. It involves the product of $\alpha_{R,\sigma}^\dagger \Rightarrow \pm(F1)D_R^\dagger d_{R,-\sigma}$ and of $c_{R',\sigma} \Rightarrow \cos\phi d_{R',\sigma}$. In the excited ("virtual") state the 2 singles disappear, as the one site is replaced by a $D$ and the other by a vacancy. The combined excitation energy is $+2V$. To leading order,

$$H_{inelastic} = t(F1)\cos\phi \times$$
$$\sum_{(R,R')} G(R-R')\{(D_R^\dagger O_{R'}^\dagger + D_{R'}^\dagger O_R^\dagger)(d_{R,\downarrow}d_{R',\uparrow} + d_{R',\downarrow}d_{R,\uparrow})(D_R D_R^\dagger)(D_{R'}D_{R'}^\dagger) + H.c.\} \quad (15)$$

is the operator that connects the low-lying initial states to the excited states. We inserted "place-markers" $O_R^\dagger$ for vacancies that are virtually created; they will be eliminated by a requirement that, in low-lying states, their occupation numbers be identically zero. We also included two factors $(D_R D_R^\dagger)(D_{R'}D_{R'}^\dagger)$ to emphasize that sites $R$, and $R'$ be initially occupied by singles only. Note that the more excess charges (i.e. $D$'s) there are, the fewer are the sites that can be occupied by singles. Next, we eliminate $H_{inelastic}$ by a unitary transformation that is basically a generalization of second-order perturbation theory.

This leaves us with an effective two-body attraction connecting pairs of singly occupied sites $PE_1$ + a hard-core repulsion among these composite particles. That is, $PE=PE_1+PE_2$, where

$$PE_1 = -\frac{(tF1\cos\phi)^2}{2V} \sum_R \sum_{R'\neq R} \sum_{R''\neq R'} G(R-R')G(R'-R'')(D_{R''}D_R^\dagger)$$
$$\times \ (d_{R',\uparrow}^\dagger d_{R'',\downarrow}^\dagger + d_{R'',\uparrow}^\dagger d_{R',\downarrow}^\dagger)(d_{R,\downarrow}d_{R',\uparrow} + d_{R',\downarrow}d_{R,\uparrow}) \cdot (D_R D_R^\dagger)(D_{R'}D_{R'}^\dagger) \quad (16)$$

This is what remains in the lowest order in $t$ after elimination of the empty sites using the condition $<O_R O_{R'}^\dagger> = \delta_{R,R'}$. The other part of $PE$, the "hard core" interaction $PE_2$, expresses the requirements that two $d$'s cannot occupy the same site $R$, that a $D$ and a $d$ cannot occupy the same site, nor can two $D$'s. It has almost been made redundant by factors $D_R D_R^\dagger = (1 - D_R^\dagger D_R)$ introduced into Eqs. (15), (16) for similar purposes.

Eqs. (14) – (16) should now be solved in *MD* or *MC* without further approximation. But that is for future investigations.

To assess the consequences we turn to a mean-field approximation. The following procedure helps construct an "effective" BCS-like Hamiltonian, with the difference that, here we need to know the statistical distribution of the charges. We use a product state for the $D$'s. This helps define a purely fermionic $H_{eff}$ by the contraction,

$$H_{eff} = \prod_R <0|(\cos\alpha_R + e^{-i\gamma_R}D_R \sin\alpha_R)(KE+PE)(\cos\alpha_R + e^{i\gamma_R}D_R^\dagger \sin\alpha_R)|0>.$$

Whatever correlations exist among the excess charges is probed using a Fourier decomposition with, as yet, unknown coefficients:

$$<D_R> = \frac{A_0 + \sum_{q\neq 0} A_q e^{-iq\cdot R}}{\sqrt{(1+\sum_q |a_q|^2)}}, \quad \text{where } a_q \equiv A_q/A_0 \quad (17)$$

Thus, $<D_{R''}D_R^\dagger> = |A_0|^2 (1+\sum_q |a_q|^2 e^{iq\cdot(R''-R)})/(1+\sum_q |a_q|^2)$ for $R''\neq R$. Similarly, $<D_R^\dagger D_R> = |A_0|^2 = <\sin^2\alpha_R> = \nu$. The symbol $\nu$ defines the "$D$ occupation number," the average excess charge *per* cell. These relations all become exact if the $\alpha_R$ are all small.

Using the inverse Fourier transform, $\sum_{R\neq 0} G(R)e^{ik\cdot R} = g(k)-G(0)$ we next exhibit the resulting Hamiltonian in the BCS pair approximation, $H_{eff} = KE_{eff} + PE_{eff}$. First,

$$PE_{eff} = -\frac{g_1\nu(1-\nu)^2}{N(1+\sum_{q\neq 0}|a_q|^2)}\sum_k\sum_{k'}\{(g(k)-G(0))(g(k')-G(0))$$

$$+\sum_{q\neq 0}|a_q|^2(g(k+q)-G(0))(g(k'+q)-G(0))\}\times(d_{k\uparrow}^\dagger d_{-k\downarrow}^\dagger)(d_{-k'\downarrow}d_{k'\uparrow})$$

$$+\frac{U^*}{N}\sum_k\sum_{k'}(d_{k\uparrow}^\dagger d_{-k\downarrow}^\dagger)(d_{-k'\downarrow}d_{k'\uparrow}) \quad , \quad \text{where } U^* = \infty. \quad (18)$$

Following a similar calculation, the motional energy becomes:

$$KE_{eff} = \frac{g_1 G(0) \nu(1-\nu)}{4} \sum_{k,\sigma} \varepsilon(k) d^\dagger_{k,\sigma} d_{k,\sigma} \qquad (19)$$

in which $\varepsilon(k) = w(k) - w_F$ is the "Bloch" single-particle energy measured from the Fermi level $w_F$. The renormalized Bloch energies are: $w(k) = \dfrac{g(k) + \sum_{q \neq 0} |a_q|^2 g(k+q)}{1 + \sum_{q \neq 0} |a_q|^2}$.

In the BCS ground state $|\Psi_0\rangle = \prod_k (\sin \frac{\theta_k}{2} + \cos \frac{\theta_k}{2} d^\dagger_{k,\uparrow} d^\dagger_{-k,\downarrow})$ the ground state energy is,

$$E_{o,pair}/N = \frac{g_1 G(0)\nu(1-\nu)}{4N} \sum_k \varepsilon(k)(\cos\theta_k + 1) - \frac{g_1 \nu(1-\nu)^2}{4N^2} \left( \sum_{k,k'} \sin\theta_k K_{k,k'} \sin\theta_{k'} \right)$$
$$+ \frac{U^*}{4N^2} \left( \sum_k \sin\theta_k \right)^2 \qquad (20A)$$

The kernel $K_{k,k'}$ in this expression, extracted from (18), is:

$$K_{k,k'} = \frac{(g(k) - G(0))(g(k') - G(0))}{(1 + \sum_{q \neq 0} |a_q|^2)} + \sum_{q \neq 0} |a_q|^2 \frac{(g(k+q) - G(0))(g(k'+q) - G(0))}{(1 + \sum_{q \neq 0} |a_q|^2)} \qquad (20B)$$

The Fermi level $w_F$, which is what determines the average number of "singles" of either spin orientation, $N - N\nu$, is obtained through a similar expression:

$$\nu = -\frac{1}{N} \sum_k \cos\theta_k \qquad (20C)$$

The $w(k)$, hence $\theta_k$, depend on the set of $\{|a_q|^2\}$. Operationally we shall want to solve for $\nu$ as a function of $w_F$, at a given set of $\{|a_q|^2\}$, then invert to obtain $w_F(\nu)$ at the given set of $\{|a_q|^2\}$ for use in Eqs. (20A) and (20B). Finally, a sort of sum rule must be obeyed:

$$\frac{1}{N}\sum_k \sin\theta_k = 0 \qquad (21)$$

If it is not, the last term in (20A) is infinite and there is no acceptable ground state in the model, not even in the mean-field approximation.[14]

*OPTIMAL CONCENTRATION OF PARTICLES.* Even before optimizing (20A) w.r. to $\sin\theta_k$ it is possible to estimate crudely, the "best" value of $\nu$. If (21) is satisfied, $E$ takes the form, $E \propto \nu(1-\nu)K - \nu(1-\nu)^2 V$, which is lowest at a density of excess carriers

$\nu_o = \dfrac{2-K/V}{3} - \sqrt{(\dfrac{2-K/V}{3})^2 - \dfrac{1-K/V}{3}}$. In many physical systems, $K/V \approx 0.5$. Using this ratio for a preliminary estimate, the formula yields $\nu_o = 0.21$, which corresponds to one extra hole for every 10.6 oxygen sites, not too far from the value of 1 hole for every 8 oxygen sites (i.e., $\nu = 0.25$) found in many optimally high-$T_c$ superconductors. But because the actual value of *K/V does* depend on $\nu$ and the $\{|a_q|^2\}$, an accurate value of $\nu_o$ can only be determined numerically after the equations for the optimal $\sin\theta_k$ are solved.

*THE GAP FUNCTIONS.* By analogy with the usual BCS hypothesis, one sets:

$$\sin\theta_k = \frac{\Delta(k)}{\sqrt{\varepsilon^2(k)+\Delta^2(k)}} = \frac{\Delta(k)}{E(k)} \quad \text{and} \quad \cos\theta_k = \frac{-\varepsilon(k)}{E(k)}, \qquad (22)$$

which also serve to define $E(k)$, the quasiparticle energy. The "gap function" or "pair wavefunction" is $\Delta(k) = \Delta_0(\varepsilon(k))y(k)$, where $\Delta_0$ is constant on a surface of constant energy. For $p$-, $d$- and higher angular momentum pairings, simple choices such as

$$y(k) = \{\cos k_x - \cos k_y\} \text{ for } d\text{-waves, and for } p\text{-waves, } y(k) = \{\sin k_x + \sin k_y\} \quad (23)$$

are adequate. The sum rule Eq. (21) is then satisfied by symmetry, even if $\Delta_0$ is constant.

The $s$-pairing equations are more complicated, albeit similar to those found in the strong-coupling limit of the Hubbard model ($t^2/U \ll 1$), *where they have no solution*.[8, 10] On the other hand, the angular $p, d, f, \ldots$ pairings are subject to no such limitations because $\frac{1}{N}\sum_k y(k)\delta(\varepsilon - \varepsilon(k))F(\varepsilon(k))$ vanishes by symmetry for arbitrary functions of the energy $F$. The gap magnitude $\Delta_0$ can be chosen to be nodeless, even constant, because symmetry eliminates the hard-core integral in $U^*$ and all constant terms in the kernel. If we now pick $q$ uniquely as $(\pi,0)$ (equivalently, $(0,\pi)$,) the ground-state energy becomes:

$$E_{o,pair}/N = \frac{g_1 v(1-v)}{4(1+|a_{(\pi,0)}|^2)}\left(S_1(\Delta_0) - (1-v)|a_{(\pi,0)}|^2 \Delta_0^2 (S_2(\Delta_0))^2\right), \quad (24)$$

where

$$S_1(\Delta_0) = (\frac{1}{2\pi})^2 \int\int_{-\pi}^{+\pi} dk_x dk_y \frac{\varepsilon(k)}{E(k)}(E(k) - \varepsilon(k)), \text{ and} \quad (25)$$

$$S_2(\Delta_0) = (\frac{1}{2\pi})^2 \int\int_{-\pi}^{+\pi} dk_x dk_y y(k) \frac{g(k+(\pi,0))}{E(k)}$$

The Fourier component at $q=(\pi,0)$[15] corresponds to (1,0) vertical or horizontal *charge stripes*.[16] These equations have to be evaluated numerically in conjunction with Eq.(20C), the equation that yields the Fermi level as a function of $v$. Clearly all integrals are functions of $w_F$ which, in turn, is a function of $v$.

Amazingly, *p*-wave pairing makes $S_2$ vanish by parity, hence is *entirely* disfavored! The *only* remaining nontrivial solutions are spin-singlet *d*-waves pairings, assumed of the form $y(k) = \{\cos k_x - \cos k_y\}$. That is what we use in the evaluation of the integral $S_2$. Plotting the energy, Eq. (24), we find a well-defined minimum developing at a finite $\Delta_0$.

*SOLVING THE MODEL.* The preceding observation leads us to minimize the energy (24) w.r. to the gap parameter $\Delta_0$ directly. This yields:

$$|a_{(\pi,0)}|^2 = \frac{1}{2(1-v)} \times \frac{I_1}{S_2 I_2} \qquad (26)$$

where $S_2$ has been defined in (25) and the two other integrals are,

$$I_1 = (\frac{1}{2\pi})^2 \int_{-\pi}^{+\pi}\!\!\int dk_x dk_y \frac{\varepsilon^2(k)y^2(k)}{E^3(k)} \quad \text{and} \quad I_2 = (\frac{1}{2\pi})^2 \int_{-\pi}^{+\pi}\!\!\int dk_x dk_y \frac{g(k+(\pi,0))y(k)\varepsilon^2(k)}{E^3(k)} \qquad (27)$$

The resulting phase diagram ($\Delta_0$, and $E_o$ vs. $v$) at a given $|a_{(\pi,0)}|^2$ will be displayed in a paper[10] under preparation. What is striking is the absence of parameters other than $|a_{(\pi,0)}|^2$. So what limits the magnitude of this effective "coupling constant" in the theory?

*ELECTROSTATICS.* The electrostatic energy associated with the shaping of charges into the form of stripes should also be taken into consideration. It is easily estimated as

$$E_{e-s}/N = +\frac{v^2}{2}(\frac{e^2}{L})\frac{|a_{(\pi,0)}|^4}{(1+|a_{(\pi,0)}|^2)^2} = g_1 \frac{v^2}{2} \frac{|a_{(\pi,0)}|^4}{(1+|a_{(\pi,0)}|^2)^2} Q. \qquad (28)$$

where $L$ is some characteristic distance. Upon setting $(e^2/L) = Qg_1$, one estimates $Q$ would have to be large, certainly $\gg 1$. As much as any other feature of our model, when $v$ is finite it is the electrostatics that will restrict $|a_{(\pi,0)}|^2$ to values O(1) or less. This aspect also is treated in the more complete work under preparation.[10]

*CONCLUSION.* The present theory is without adjustable parameters, except for $g_1=t^2/V$ that sets the over-all scale of energies; $V$ is large and $U=\infty$. Each cell has well-defined states of 0 ("vacancy"), 1 or 2 particles.[10] For particle concentrations *per* cell exceeding 1 on average, "vacancy" states are pretty well excluded by the energetics. The motility of the excess charges depends very much on their concentrations. But any two sites, hosting a single particle each, scatter inelastically into one doubly-occupied site + a vacancy. The vacancies are eliminated by a canonical transformation, generating a two-body attraction among singly-occupied sites, plus a hard core. This translates into Cooper pairing, although the hard core inhibits *s*-wave pairs. But if the excess charges are correlated into stripes, *d*-wave pairs are promoted. The only undetermined parameter is the strength of the stripe Fourier component $|a_q|^2$. Given $|a_q|^2$, the lowest energy is found at $v \approx 20\%$. We predict that a stripe oriented at 45º (i.e. $q=(\pi,\pi)$,) *cannot* promote superconductivity whereas $q=(\pi,0)$ does. Moreover, once electrostatic forces are included the intensity of the stripe $|a_{(\pi,0)}|^2$ can actually be determined, *leaving no more parameters to adjust*.

*ACKNOWLEDGEMENTS.* This research was supported by a University of Utah seed grant No. 51003084 (2004-2005) and by NSF grant No. ECS-0524728 (2005–7), which are gratefully acknowledged.

FOOTNOTES & REFERENCES

---

[1] introduced in the early 1960's by John Hubbard, Proc. Roy. Soc. **A276**, 238 (1963), M.C. Gutzwiller, Phys. Rev. Lett. **10**, 159 (1963) and Phys. Rev. **137**, 1726 (1965)

[2] At the heart of the Hubbard model lies the two-body repulsion *U* of two electrons of opposite spin on the same site, together with complete screening for particles localized on distinct sites.

[3] E.g., Th. Maier, M. Jarrell, Th. Pruschke and J. Keller, Phys. Rev. Lett. **85**, 1524 (2000), see also D. Eichenberger and D. Baeriswyl, in ArXiv: cond-mat/0608210v1 (9 Aug. 2006)

[4] G. Migliorini and A.N. Berker, Eur. J. Phys. **B17**, 3 (2000)

[5] E. Dagotto, Rev. Mod. Phys. **66**, 763–840 (1994). Dagotto's paper reviewed several extant theories, including Hubbard and *t-J* models. In summary, while the Hubbard model is compatible only with *s*-wave gaps (if any– *cf.* refs. 5 below), the *t-J* model *is* compatible with the *d*-wave gap that is observed in high $T_c$ superconductivity. However, as a physical model, it is incomplete (this was originally shown by Hirsch in ref. 8 below.)

[6] Consider the analysis by C.-N. Yang. Phys. Rev. Lett. **63**, 2144 (1989) and Shoucheng Zhang, Phys. Rev. **B42**, 1012 (1990), who showed that *if* pairing were favored the pairing amplitude (gap function) would have to be an *s*-wave, i.e. $\Delta(k)$ would exhibit radial nodes in *k*-space while maintaining full $C_4$ symmetry.

[7] E. H. Lieb and F.Y. Wu, Phys. Rev. Lett. **20**, 1445 (1968) and Physica **A321**, 1 (2003) solved the Hubbard model definitively in 1D, where it shows absolutely *no* propensity whatever for superconductivity. On the contrary, antiferromagnetism *is* preferred at half-filling and paramagnetism (not diamagnetism) at all other concentrations.

[8] J. Hirsch, Phys. Rev. Lett. **54**, 1317 (1985): his strong-coupling reduction compares well to exact results on the Hubbard model in 1D or in small clusters in higher dimensions. It contains, in addition to the terms in the *t-J* model, a number of 3-center terms. If it had a superconducting solution, it could only be *s*-wave (in contradiction to experiment.) The *t-J* model, which *does* allow for *d*-wave pairings, results only when Hirsch's three-center terms are arbitrarily omitted and therefore cannot be trusted as a microscopic theory of high-$T_c$ superconductivity. The addition of stripes to the Hubbard model does allow *d*-wave pairings to form. However, these are squelched in strong-coupling by the small value of the coupling constant, $t^2/U$.

[9] One obsolete version (without large *V*) is by the present author, D.C. Mattis, Phys. Rev. Lett. **74**, 3676 (1994) and Mod. Phys. Lett. **B21**, 1387 (1994). Other treatments include G. Dopf, A. Muramatsu and W. Hanke, Phys. Rev. Lett. **68**, 353 (1992), J.H. Jefferson, H. Eskes and L.F. Feiner, Phys. Rev. **B45**, 7959 (1992), L.F. Feiner, J.H. Jefferson and R. Raimondi, Phys. Rev. **B53**, 8751 (1996). There exist many more such tight-binding models of oxides and compounds. One can trace their ancestry to J. Zaanen, G.A. Sawatzky and J.W. Allen, Phys. Rev. Lett. **55**, 418

(1985), V. J. Emery, Phys. Rev. Lett. **58**, 2794 (1987) and C. M. Varma, S. Schmitt-Rink and E. Abrahams, Solid St. Commun. **62**, 550 (1987)

[10] N.B.: the particles in this theory are, and will consistently be, *holes*. When $U$ is in the range $2V < U < \infty$, a parameter $g_2 = t^2/U$ appears, multiplying additional, rather complex, operators in the Hamiltonian. The theory becomes slightly modified, but our conclusions are not. This is shown in a paper under preparation: D.C. Mattis *et al*, to be submitted

[11] The $\beta$'s are out of the picture if we restrict the Hilbert space to the set of low-lying basis states of energy in the range $-V < E < 0$.

[12] Inclusion of triplet pairs raises the number of states from 4 at each site to 7. Although the final results are affected only slightly (channels separate by symmetry and triplets become excluded), the notation required to include them becomes seriously complicated. For lack of space we shall omit further discussion here but refer to the analysis in ref. 10.

[13] That this insulator is also antiferromagnetic is a feature that can only appear in *fourth* order in $t$, as well known in the theory of superexchange. For present purposes it is unnecessary to go beyond *second* order in $t$. Additionally, for occupation numbers below 1 particle *per* site, it is required to reintroduce *empty* sites, while it is the $D$'s that become unwelcome in the low-lying eigenstates. This topic also is examined in ref. 10.

[14] There are two ways to ensure this: by a function $\theta_k$ having one or more nodes as $\varepsilon(k)$ is increased (*s*-pairing,) or having one or more nodes along each contour of constant $\varepsilon(k)$ (denoted *p, d, f, ...* wave pairings.)

[15] At given $|a_q|^2$ the integral $S_2$ increases with $|q|$. Thus $q=(\pi,0)$ is the best choice *a priori* corresponding to vertical or horizontal striping, at twice the lattice parameter (this would be $\approx 7.5$ Å in $CuO_2$). At smaller $q$'s one must also include $-q$. Interestingly, what might otherwise have been a competing choice, $q = (\pi,\pi)$, yields $S_2 \equiv 0$ *by symmetry*. It is therefore not at all conducive to *any* pairing, whether *s-, p-* or *d*-wave. This prediction is, of course subject to future experimental verification.

[16] The whole subject of stripes, while new to this author, arose naturally in his attempt to build a superconducting ground state for the Hamiltonian in Eqs. (14)-(16). In retrospect, striped structures had already appeared inevitable to pioneers in this field, some of whom have discussed several types: spin and charged stripes which appear in mosaic, checkerboard and other lower symmetry structures that we have not had the leisure to investigate. Spin stripes play no apparent

role in the present theory, perhaps because the "superexchange" parameter $J$ is $O(t^4)$ and we calculate only to $O(t^2)$. A zero-field unambiguous experimental observation of stripes was recently reported by C. Howald, H. Eisaki, N. Kaneko and A. Kapitulnik, Proc. Nat. Acad. Sci. (USA) **100**, 9705 (2003). Striped phases were reported by J. M. Tranquada, B. J. Sternlieb, J.D. Axe, Y. Nakamura and S. Uchida, Nature **375**, 561 (1995), following general predictions by V.J. Emery, S.A. Kivelson and H.Q. Lin, Phys. Rev. Lett. **64**, 475 (1990) and V.J. Emery and S.A. Kivelson, Physica **C209**, 597 (1993). Useful reviews (with many additional references) include A. M. Olés, Acta Physica Polonica **B31**, 2963 (2000) and S.A. Kivelson, E. Fradkin, V. Oganesyan, J.M. Tranquada, A. Kapitulnik and C. Howald, Rev. Mod Phys. **75**, 1201-1241 (2003).